УДК 378:004.4'236

**В. Є. Величко,**
кандидат фізико-математичних наук, доцент, докторант
(Луганський національний університет імені Тараса Шевченка)
vladislav.velichko@gmail.com

# ВИКОРИСТАННЯ ТЕХНОЛОГІЇ ВІЗУАЛЬНОГО ПРОГРАМУВАННЯ В УНІВЕРСИТЕТСЬКІЙ ОСВІТІ ЗАСОБАМИ ВІЛЬНОГО ПРОГРАМНОГО ЗАБЕЗПЕЧЕННЯ

*Термін "візуальне програмування" увійшло до термінології з інформатики досить давно, тим не менш існують різні погляди на його значення. Відокремлення візуального програмування від засобів розробки інтерфейсів надає не тільки визначеності цій технології, а й відкриває для освіти невідомі широкому колу системи моделювання та програмування. Використання технології візуального програмування, яка основана на вільному програмному забезпеченні, надає нові перспективи з формування алгоритмічної культури майбутніх фахівців некомп'ютерних спеціальностей. Широке впровадження візуального програмування в освітній процес є перспективним напрямом інтенсифікації пізнавальної діяльності, закладенням основ алгоритмічної компетентності.*

***Ключові слова:*** *візуальне програмування, вільне програмне забезпечення, алгоритмічна культура.*

**Постановка проблеми в загальному вигляді.** Інформаційні технології, що широко використовуються в освіті, охоплюють широке коло питань. Різноплановість програмно-апаратних засобів освітньої діяльності дозволяє їх використання для підготовки майбутніх фахівців будь-якого напрямку – технічного, гуманітарного, інженерного, філософського тощо. А отже, питання підготовки як суб'єктів, так і об'єктів навчання до використання інформаційних технологій виходить на одне з перших місць серед завдань вищої освіти. Однією із цілей Закону України "Про Основні засади розвитку інформаційного суспільства в Україні на 2007-2015 роки" є забезпечення комп'ютерної та інформаційної грамотності населення, насамперед, шляхом створення системи освіти, орієнтованої на використання новітніх інформаційно-комунікаційних технологій у формуванні всебічно розвиненої особистості.

Перехід до користувацького напрямку підготовки в освітньому напрямку "Комп'ютерні науки" видозмінили не тільки цілі та методи навчання, але й поставили нові завдання до підготовки фахівців. Однак, до повноцінної комп'ютерної та інформаційної грамотності варто віднести також і алгоритмічну культуру, яка характеризується усвідомленням значимості процесу алгоритмізації, певним рівнем алгоритмічного мислення й проявляється в різноманітних формах і способах організації та самоорганізації алгоритмічної діяльності. Таким чином, необхідно приділити увагу розвитку алгоритмічної культури майбутніх фахівців будь-якого напрямку навчання з врахуванням сучасного розвитку інформаційно-комунікаційних технологій.

**Аналіз останніх досліджень і публікацій.** У психолого-педагогічній літературі накопичено достатню кількість досліджень, пов'язаних із застосуванням інформаційно-комунікаційних технологій в освіті. Підходи до комп'ютеризації навчального процесу як в загальному вигляді, так і в певних напрямках розглянуто в працях В. Бикова, Б. Гершунського, А. Єршова, М. Жалдака, В. Ізвозчикова, Ч. Кларка, К. Коліна, М. Лапчика, Ю. Машбиця, В. Монахова, І. Підласного, Є. Полат, Ю. Рамського, І. Роберт, Д. Севедж, Г. Селевка, О. Спіріна, Н. Тализіної, Ю. Тріуса та ін. Дослідженню використання інформаційно-комунікаційних технологій у новітніх формах організації навчального процесу присвячені дослідження О. Андреєва, С. Архангельського, Т. Гусакової, Н. Кузнецової, В. Кухаренка, Н. Морзе, В. Олійника, Є. Полат, П. Стефаненка, П. Таланчука, А. Хуторського, Б. Шуневича та ін. Загальні проблеми вільного програмного забезпечення, юридичні та філософські аспекти його існування та використання висвітлюються в роботах Дж. Гослінга, Е. Реймонда, Р. Столлмана та ін., використання вільного програмного забезпечення в системі освіти присвятили свої роботи В. Габрусєв, О. Дима, М. Карпенко, М. Кияк, О. Нестеренко, Л. Панченко, А. Панчук, В. Хахановський.

**Мета дослідження** полягає в аналізі можливостей використання технології візуального програмування в університетській освіті засобами вільного програмного забезпечення.

**Виклад основного матеріалу.** Формування алгоритмічної культури складається з наступних компонентів: розуміння сутності алгоритму і його властивостей, розуміння сутності мови запису алгоритму, володіння навичками і засобами запису алгоритму, розуміння алгоритмічного характеру наукових методів і їх використання. Актуальність алгоритмічної культури не означає, що необхідно повернутись до попередньої концепції навчання предметам циклу "Інформатика", але, для висококваліфікованого фахівця необхідно сформувати, як мінімум, базові навички.

Сучасні парадигми програмування досить складні й багатогранні, не кажучи вже про їх реалізації. Окрім навчальних мов програмування, існують і спеціальні, призначені для вирішення певного кола задач. Однак, загальні принципи та методи програмування можна прослідкувати, розглядаючи візуальне програмування.

**© Величко В. Є., 2014**

Історія розвитку візуальних мов програмування почалась з кінця 70-х – початку 80-х років. Основна ідея, яка покладена в основу візуального програмування, полягає в тому, що візуальна мова програмування – це мова, яка використовує деякі візуальні представлення (в додаток або замість слів та чисел) для обчислення того, що в протилежному випадку було б виражено в традиційні формі [1]. В той же час у підручниках, довідниках і словниках з віртуального програмування особливо підкреслюється, що дуже популярні системи програмування типу MS Visual Studio, Delphi, які здійснюють супровід процесу об'єктно-орієнтованого проектування та програмування, не можуть претендувати на звання класичних систем візуального програмування, незважаючи на розповсюдженість такого найменування [2: 42].

Сучасний етап розвитку технологій візуального програмування характеризується підвищеним попитом на дані технології. Системи моделювання та аналізу здатні виконувати завдання зі швидкої розробки програмного забезпечення. Широке розповсюдження мобільних сенсорних пристроїв та їх перспективне використання в навчанні, зокрема в мобільному навчанні, спонукають до розробки навчальних засобів і візуального програмування [3].

Візуальне програмування – спосіб створення програм шляхом маніпулювання графічними об'єктами замість написання програмного коду в текстовому вигляді. Варто розрізняти різні види візуального програмування:

- графічна мова програмування (мова зі своїм синтаксисом);
- візуальні засоби розробки (засоби проектування інтерфейсів, CASE-системи швидкої розробки додатків, SCADA-система для програмування мікроконтролерів тощо).

Візуальне програмування тісно пов'язане з поняттям моделювання, що використовується в парадигмі об'єктно-орієнтованого програмування та є невід'ємною частиною уніфікованого процесу розробки програмного забезпечення. Найпопулярнішим відкритим стандартом візуального моделювання є стандарт UML (*Unified Modeling Language*), створений для визначення, візуалізації, проектування й документування в основному програмних систем. Він не є мовою програмування, але в засобах виконання UML-моделей як інтерпретованого коду можлива кодогенерація [4]. Візуалізація моделювання при всіх своїх перевагах не позбавлена недоліків і скептичного ставлення до неї. Піддаються сумнівам не тільки реалізації візуального моделювання, а й сама ідея доцільності такого моделювання. Водночас, існує великий клас програмного забезпечення, який, використовуючи візуальне моделювання, дозволяє не тільки будувати динамічні моделі, а й досліджувати їх. Для прикладу, середовище візуального проектування Xcos (SciCos), яке є додатком до вільно розповсюджуваного пакету прикладних програм Scilab, дозволяє створювати та аналізувати моделі динамічних систем [5]. Доволі корисним продуктом є розробка в цьому напрямку дослідників фірми "Експериментальні об'єктні технології" при Санкт-Петербурзькому державному політехнічному університеті Model Vision Studium, яка на жаль переросла до пропрієтарної системи AniLogic. Ці й не тільки ці програмні системи широко використовуються в університетській освіті, про що говорить значна кількість виконаних досліджень.

Наступним класом вільного програмного забезпечення є системи візуального програмування, основною метою яких є підвищення розуміння графічного представлення алгоритмів для людського зорового сприйняття. Деякі вчені вважають, що існуючі способи запису алгоритмів і програм занадто важкі для розуміння й вимагають невиправдано великих трудовитрат. Ця обставина ставить нездоланний бар'єр для не програмістів, тобто фахівців, робота яких пов'язана з алгоритмами, але які не мають резерву часу, щоб навчитися виражати свої професійні знання у формі алгоритмів і програм [6]. Позиціонуєма як "доброзичлива російська алгоритмічна мова, яка забезпечує наочність ДРАКОН", використовує нову ергономічну нотацію (дракон-схеми) й за рахунок цього суттєво полегшує алгоритмізацію та програмування. На думку фахівців, завдяки використанню дракон-схем алгоритми та програми стають більш зрозумілими, дохідливими, ясними, прозорими [7].

Середовище візуального програмування Scratch від MIT побудовано на принципах мови Logo та реалізує концепцію дитячого конструктора Lego, дозволяє візуально, оперуючи об'єктами середовища створювати алгоритми керування виконавцями. І хоча дане середовище позиціонується для молодшого шкільного віку, візуалізація не тільки процесу створення алгоритму, але і його виконання дозволяє студентам нематематичних спеціальностей формувати основні елементи алгоритмічної компетентності. Середовище стає інструментом організації розумової діяльності, формуванням культури конструктивного пізнання, в основі якої ідеї – самоорганізації, саморозвитку, самореалізації особистості, що відповідають цілям компетентнісної освіти, покладеної в основу освітніх стандартів нового покоління [8].

Продовженням ідей середовища Scratch є реалізований мовою JavaScript засіб програмування Google Blocky. Результат візуального складання блоків в Google Blocky можна скомпілювати в одну з традиційних мов програмування JavaScript, Dart або Python. Каліфорнійський університет в Берклі надає за відкритою ліцензією своє середовище Snap (Build Your Own Blocks), яке відповідає сучасним стандартам HTML5/JavaScript. Аналогічна розробка існує також і для платформи Mac OS X – це мова візуального програмування Stencyl, яка спеціалізується на швидкому створенні ігрових додатків для iOS та Flash. Не залишилась осторонь і платформа Android, для якої створено середовище візуального програмування App

Inventor, із компіляцією візуальної блочної мови в байт-код Andoid. Система з назвою СтройКод дозволяє створювати не тільки виконувані файли для платформи Windows, а й Web-додатки та програми й модулі в Delphi та Free Pascal. Не кожна із наведених систем має прикладне значення, але використання їх в навчальній діяльності дозволяє формувати алгоритмічну культуру студентів, давати знання про властивості алгоритмів та способи розв'язання задач за допомогою алгоритмів.

Програмування анімованого 3D-середовища доступне в засобі розробки візуальної об'єктно-орієнтованої мови програмування Alice. Однією із задач, що ставлять перед собою автори – направленість на певний прошарок населення, який не використовує в своїй діяльності класичне комп'ютерне програмування. Середовище Alice дозволяє створювати так звані "віртуальні світи", використовуючи величезний набір персонажів, об'єктів тощо. Подійно-орієнтоване програмування засобами візуального створення алгоритмів дії привносить в створений віртуальний світ певні прогнозовані дії персонажів. Досліди, що проводили вчені, дали позитивний результат із використання цієї системи на студентах некомп'ютерних спеціальностей не тільки підвищенням рівня їх оцінок з напряму "Комп'ютерні науки", а й підвищенням рівня засвоєння інформації з 47 % до 88 % [9].

Наступним класом візуального програмного забезпечення є конструктори програм, тобто такі середовища, в яких користувач буде програму, додаючи елементи інтерфейсу, елементи контролю, елементи перетворення даних тощо та створюючи зв'язки між ними. Прикладом такого середовища є HiAsm – система, що дозволяє створювати додатки мовою Delphi та Free Pascal для платформ Android та Windows CE, Web-додатків з використанням PHP, JavaScript. Так би мовити платою за використання візуалізації процесу програмування є обмеження всіх можливостей об'єктів середовища до найбільш уживаних. Середовище HiAsm не призначене для використання в навчальній діяльності, але, для створення повноцінних програмних продуктів тими студентами, що мають певний досвід програмування, і може слугувати додатковим засобом розробки програмного забезпечення.

За загальноприйнятою класифікацією вільне програмне забезпечення, яке підтримує технологію візуального програмування, відносять за ступенем взаємодії з апаратною частиною обчислювальної системи до:
- прикладного програмного забезпечення;
- інструментальних засобів програмування;
- засобів візуалізації.

За ліцензійністю його відносять до:
- вільного програмного забезпечення (GNU GPL, Apache License, MIT License, BSD License);
- суспільного надбання (Public Domain).

**Висновки та перспективи подальших розвідок.** Візуальне програмування є сучасною технологією, що не тільки досягла певних успіхів, а й має напрямки та перспективи розвитку. Хоча вона поки й не набула промислових масштабів у програмуванні, освітній галузі, дана технологія вже має досвід застосування. Результати використання візуального програмування у формуванні алгоритмічної культури незаперечні, а отже подальше використання із вдосконаленням методик дозволить розвинути алгоритмічну компетентність майбутніх студентів.

Проаналізувавши вільне програмне забезпечення з технологією візуального програмування, можна виділити унікальні можливості систем і засобів візуального програмування, реалізація яких створює передумови для інтенсифікації освітнього процесу, а також створення методик, що орієнтовані на розвиток особистості:
- миттєвий обернений зв'язок між користувачем та моделлю, алгоритмом;
- комп'ютерна візуалізація навчальної інформації про об'єкти або закономірні процеси, явища, що реально відбуваються або віртуально;
- збереження напрацьованої інформації відносно моделей та алгоритмів із можливістю її передачі, а також організації загального доступу до неї;
- автоматизація процесів обчислювальної, інформаційно-пошукової діяльності, в тому числі й обробка результатів експериментів із можливістю багатократного повторення фрагменту або самого експерименту.

Реалізація вищенаведених можливостей дозволяє організувати такі види навчальної діяльності, як:
- реєстрація, збір, накопичення, збереження, обробка інформації про об'єкти, явища, процеси, які вивчаються або досліджуються, та передача інформації, яка представлена в різноманітних форматах;
- інтерактивний візуальний діалог взаємодії користувача із системою, що характеризується на відміну від звичайного діалогу візуальними нетекстовими образами;
- керування реальними об'єктами, спеціалізованими засобами моделювання діяльності;
- керування зображенням на екрані моделей різноманітних об'єктів, явищ, процесів, в тому числі й тих, які реально відбуваються;
- автоматизований контроль (самоконтроль) результатів навчальної діяльності, корекція за результатом контролю, тренування, тестування.

***Величко В. Е. Использование технологии визуального программирования в университетском образовании средствами свободного программного обеспечения.***

*Термин "визуальное программирование" вошло в терминологию по информатике достаточно давно, тем не менее существуют различные взгляды на его значение. Отделение визуального программирования от средств разработки интерфейсов предоставляет не только определенность этой технологии, но и открывает для образования неизвестные широкому кругу системы моделирования и программирования. Использование технологии визуального программирования, основанной на свободном программном обеспечении, предоставляет новые перспективы по формированию алгоритмической культуры будущих специалистов некомпьютерных специальностей. Широкое внедрение визуального программирования в образовательный процесс является перспективным направлением интенсификации познавательной деятельности, фундаментом алгоритмической компетентности.*

***Ключевые слова:*** *визуальное программирование, свободное программное обеспечение, алгоритмическая культура.*

***Velichko V. E. The Use of Visual Programming in the University Education by Means of Free Software.***

*The term "visual programming" has started to be used in Informatics so far, however, there are different views on its meaning. The separation of visual programming from development tools of interfaces provides not only the certainty for this technology, but also opens the systems of modeling and programming in the education unknown to a wide range. The use of visual programming which is based on the free software provides the new perspective algorithmic culture formation of future professionals of non-computer specialists. A wide application of visual programming in the educational process is the perspective direction of the cognitive activity intensification, laying the foundations of the algorithmic competence. Visual simulation systems which are related to the creation of the unified means of any software, when are used in training activities allow not only to create models, determining the nature, categories and connections, but also to perform the experimental calculation of the model's conduct for parametric input data, identify behaviour of the system. Therefore the detection problem of the free software that supports visual programming is relevant and for its solution the general features of such systems and ways of their use have been analyzed. These results suggest the significant potential use of visual programming technologies in the educational activities and that there exist the free software which implements this technology and only enhances its value. Visual programming is an advanced technology, but also has directions and prospects. Although it has not got a large scale in programming yet but it has the experience in the educational application. The results of the visual programming use in the algorithmic culture formation are undeniable, and thus the further use with improvement techniques will allow to develop the future students' algorithmic competence.*

***Keywords:*** *visual programming, free software, algorithmic culture.*